%% file: main.tex
\documentclass[hidelinks]{IEEEtran}
\usepackage{amsmath,amssymb,amsfonts}
\usepackage{graphicx}
\usepackage{textcomp,nicefrac}
\usepackage{xurl}
\usepackage{doi}

\usepackage{tabularx}
\usepackage{booktabs}
\usepackage{multirow}

\usepackage{listings}

\usepackage{fancyvrb}
\usepackage{relsize}
\usepackage{hyperref}
\usepackage{comment}

\usepackage{acro}

\input{acronyms.tex}

\def\BibTeX{{\rm B\kern-.05em{\sc i\kern-.025em b}\kern-.08emT\kern-.1667em\lower.7ex\hbox{E}\kern-.125emX}}

\begin{document}
\title{Cross-Chip Partial Reconfiguration for the Initialisation of Modular and Scalable Heterogeneous Systems}
\author{Marvin Fuchs, Hendrik Krause, Timo Muscheid, Lukas Scheller, Luis E. Ardila-Perez, Oliver Sander
\thanks{Manuscript submitted April 8, 2024; revised June 21, 2024.

This work was funded by the Federal Ministry of Education and Research (BMBF) within the framework programme "Quantum technologies – from basic research to market" (Project QSolid, Grant No. 13N16151 and Project QBriqs, Grant No. 13N15950).
This research acknowledges the support by the Doctoral School\emph{``Karlsruhe School of Elementary and Astroparticle Physics: Science and Technology''}.

M. Fuchs (corresponding author, email: marvin.fuchs at kit.edu),
H. Krause, T. Muscheid, L. Scheller, L. E. Ardila-Perez and O. Sander are with the Institute for
Data Processing and Electronics (IPE) of the Karlsruhe Institute of Technology,
Hermann-von-Helmholtz-Platz 1, D-76344 Eggenstein-Leopoldshafen, Germany}}


\maketitle

\begin{abstract}

The almost unlimited possibilities to customize the logic in an \acs{fpga} are one of the main reasons for the versatility of these devices. \Acl{pr} exploits this capability even further by allowing to replace logic in predefined \acs{fpga} regions at runtime. This is especially relevant in heterogeneous \acsp{soc}, combining \acs{fpga} fabric with conventional processors on a single die. Tight integration and supporting frameworks like the \acs{fpga} subsystem in Linux facilitate use, for example, to dynamically load custom hardware accelerators. Although this example is one of the most common use cases for \acl{pr}, the possible applications go far beyond. We propose to use \acl{pr} in combination with the \acs{axi} \acs{c2c} cross-chip bus to extend the resources of heterogeneous \acs{mpsoc} and \acs{rfsoc} devices by connecting peripheral \acsp{fpga}. With \acs{axi} \acs{c2c} it is easily possible to link the programmable logic of the individual devices, but partial reconfiguration on peripheral \acsp{fpga} utilising the same channel is not officially supported. By using an \acs{axi} \acs{icap} controller in combination with custom Linux drivers, we show that it is possible to enable the \acs{ps} of the heterogeneous \acs{soc} to perform \acl{pr} on peripheral \acsp{fpga}, and thus to seamlessly access and manage the entire multi-device system. As a result, software and \acs{fpga} firmware updates can be applied to the entire system at runtime, and peripheral \acsp{fpga} can be added and removed during operation.

\end{abstract}

\begin{IEEEkeywords}
Partial Reconfiguration, Dynamic Function Exchange, DFX, AXI, MPSoC, System-on-Chip, Zynq UltraScale+
\end{IEEEkeywords}

\section{Introduction}
\label{sec:introduction}

\IEEEPARstart{P}{artial reconfiguration} is a feature of \acfp{fpga} that enables exchanging the logic in one part of the \ac{fpga} at runtime without affecting the operation of the remaining parts. It was recognized decades ago that this feature can expand the application possibilities of \acp{fpga} \cite{Mignolet2003, Becker2003}. For instance, because multiplexing the logic in the device at runtime allows for a more efficient usage of the available hardware resources. In applications where this is possible, a smaller device can be selected, which does not only reduce costs but also power consumption. Although \acp{fpga} already supported \acl{pr} at that time, the feature was rarely employed in practice because it could not always be used reliably and entailed a lot of additional development effort. This has changed with the improvement of the development tools and especially with the introduction of heterogeneous \acf{soc} devices, combining \ac{fpga} fabric with conventional processors on the same chip. In such an \ac{soc}, there are significantly more use cases for \acl{pr}, as the \acf{ps} can be used to manage the partial bitstreams and to control the \acl{pr}. One example is the PYNQ framework from AMD Xilinx \cite{pynq}. It provides a high-level Python \ac{api} for the \ac{ps} that allows the user to easily interact with the \ac{soc}'s \ac{pl} and to reconfigure all or predefined parts of it with a single command. In this way, PYNQ supports the usage of heterogeneous \acp{soc} as an adaptive computing platform by allowing hardware accelerators to be loaded and used intuitively.

In this work, the possibilities offered by the \ac{ps} were extended even further, whereby not only the \ac{pl} within the same chip can be partially reconfigured, but also external \acp{fpga} connected via \acf{axi} \acf{c2c}. \Ac{axi} \ac{c2c} is a method provided by AMD Xilinx to connect the internal \ac{axi} busses of two devices via \acp{gt}. Being able to perform \acl{pr} on external devices is an important step towards being able to seamlessly expand the resources of an heterogeneous \ac{soc} by connecting peripheral \acp{fpga}. The expansion of resources can be useful in various cases, but our work is primarily motivated by the needs of modern \ac{daq} systems for physics experiments.

\subsection{Heterogeneous SoCs in DAQ Systems}

\Ac{daq} systems for physics experiments must push the limits of the latest available technology to fully exploit the potential of novel detectors. At the same time, these systems must be adapted to meet the specific requirements of the particular experiment. Heterogeneous \acp{soc} are often a good platform to meet these requirements, because the \ac{ps} can handle complex control schemes and allow to seamlessly integrate the device into existing Ethernet networks, whereas the \ac{pl} allows to implement application-specific, fast real-time modules and custom interfaces. To support different applications, AMD Xilinx provides two families of heterogeneous \acp{soc}. \Acfp{mpsoc} contain only the \ac{ps} and the \ac{pl}, whereas \acfp{rfsoc} additionally contain \acp{adc} and \acp{dac}. As the following examples show, the advantages of heterogeneous \acp{soc} are beneficial from large, distributed setups with multiple hundreds of \acp{soc} and \acp{fpga} down to small single-device systems.

Various custom electronics cards are being developed on the basis of the \ac{atca} specification for the next upgrade of both general-purpose large-scale experiments ATLAS and CMS at CERN \cite{Mehner2024, Mehner2022, Albert2022, Tang2022, Loukas2022}. Most of them will use the versatility of heterogeneous \acp{mpsoc} from AMD Xilinx and combine them with one or more large \acp{fpga} on an \ac{atca} blades to form a unit with more resources. However, to meet the requirements of the experiments, hundreds of these cards will be combined to form a massive \ac{daq} system.

This contrasts, for instance, with systems to interface cryogenic sensors, based on a single \ac{mpsoc} or \ac{rfsoc} \cite{Muscheid2023, Smith2022, Gebauer2022_thesis}. One such control and \acs{daq} system is the so-called ``QiController''.

\subsection{QiController}

\begin{figure}[t]
    \centering
    \includegraphics[width=3.4in]{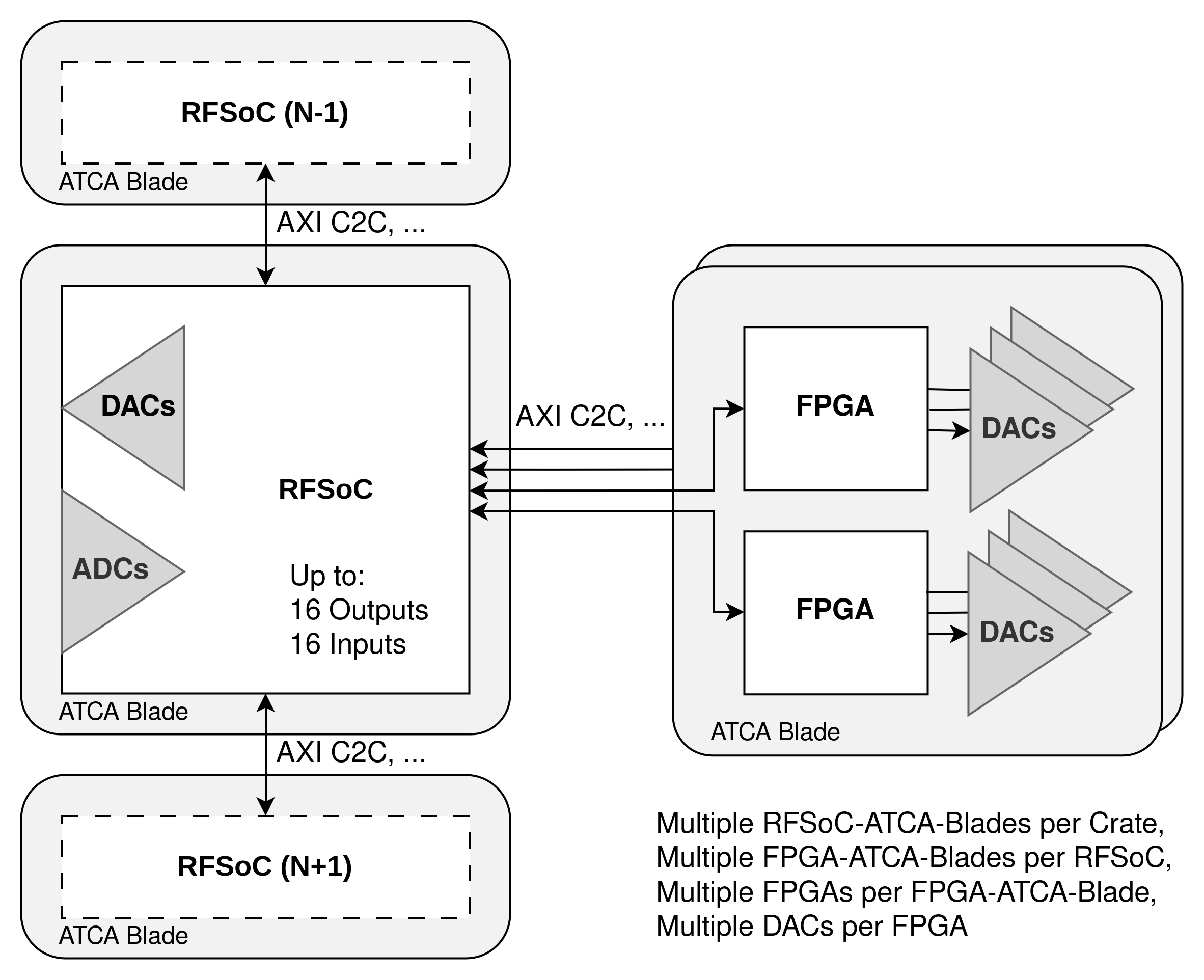}
    \caption{Proposed modular and scalable multi-device QiController architecture. All connections between \ac{atca} blades are established via the \ac{atca} backplane.}
    \label{fig:multiBoardPlatform}
\end{figure}

The QiController is a \ac{sdr} system primarily dedicated to the characterization of superconducting qubits \cite{Krantz2019}. Currently, it is based on a single \ac{rfsoc} from AMD Xilinx with sixteen \ac{adc} and sixteen \ac{dac} channels. To extend the range of applications and increase the number of addressable qubits, the system is currently being expanded to a hardware platform composed of multiple \acp{rfsoc} that are interconnected to several peripheral \acp{fpga} and to one another. The proposed architecture is depicted in \autoref{fig:multiBoardPlatform}. In this configuration, the analogue channels of the \acp{rfsoc} are directly used to interface with the qubits, while the peripheral \acp{fpga} are used to drive further \acp{dac} that increase the number of analogue channels and interface the qubit coupling devices. This allocation was chosen because the frequency and resolution requirements for qubits and coupling devices are different.

Expanding the single-chip-based QiController to a multi \ac{soc} and multi \ac{fpga} platform introduces a variety of new challenges, including the proper partitioning of the system into several electronic cards, a multi-device initialisation sequence that takes into account the specific requirements of each individual component, and an intuitive and easy-to-use method for updating the software and \ac{fpga} firmware on the system. To make the hardware modular and expandable we chose to design it according to the \ac{atca} specification. However, the \ac{atca} backplane limits the number of connections between individual cards to only four bi-directional lanes. One of these is used to connect the \ac{pl} of an \ac{rfsoc} to a peripheral \ac{fpga} via \ac{axi} \ac{c2c} in order to link the logic in both devices. We propose to use precisely this connection to perform \acl{pr}, which allows the \ac{fpga} firmware on the peripheral \ac{fpga} to be initialised, updated, and dynamically exchanged from the processors in the \ac{rfsoc}, efficiently using the available connections between devices.

\section{Related Work}

Since a few years, large heterogeneous systems utilising various computational engines, including \acp{fpga}, \acp{cpu}, \acp{gpu}, and \ac{ai} specific processors, are frequently deployed in data centres of major providers like Alibaba, Amazon, Baidu, Huawei, and Microsoft \cite{Bobda2022}. This trend is motivated by the expectation that specialised processing units offer the biggest potential for hardware-based computational performance gains. Mainly for this reason, large heterogeneous systems with multiple \acp{fpga} are the subject of current research \cite{Ammendola2017, Flich2021, Ioannou2020}. The automatic and fast \mbox{(re-)configuration} of \acp{fpga} is a crucial part of this research, as it is a special requirement of these devices that other computational engines do not have, making it a new challenges for a data centre scenario. One example is that during \mbox{(re-)configuration} the \ac{fpga} cannot be used, which means a non-negligible dead time. Furthermore, unwanted loading of configuration data can lead to potential security risks. However, solutions for initialising and reconfiguring \acp{fpga} in data centres, which are commonly based on \ac{pcie} \cite{aws_faq, ruan2022} or Ethernet \cite{pham2020}, are tailored to the special requirements of that field and cannot be universally applied to \acs{daq} systems. There are various reasons for this, most of which are a result of the different requirements made on the systems. Data centers aim to achieve best possible computational performance, whereas \acs{daq} systems have real time requirements and must ensure integrity of the measurement data. This can also be seen in the fact that heterogeneous systems in data centres are usually administered with many layers of software, allowing for automated management and steering of the multi-device setup process and data flow. In \acs{daq} systems, it is typically preferred to avoid this level of overhead and uncertainty by instead using a leaner, customized solution. Nonetheless, there are mechanisms implemented in data centres that can be transferred to the \ac{daq} domain, such as \ac{fpga} programming via \acl{pr} as used by \ac{aws} \cite{aws_faq}. This concept includes a small static framework that is automatically loaded on the \ac{fpga} at power-up to initialise various of its interfaces, which can later be used to load the payload logic using \acl{pr}. The mechanism presented here also employs this method, but in combination with the \ac{axi} protocol, which is a frequently used option for intercommunication in heterogeneous system consisting of heterogeneous \acp{soc} and \acsp{fpga}.

\section{Expanding an Heterogeneous \ac{soc}}

\begin{figure}[b]
    \centering
    \includegraphics[width=3.4in]{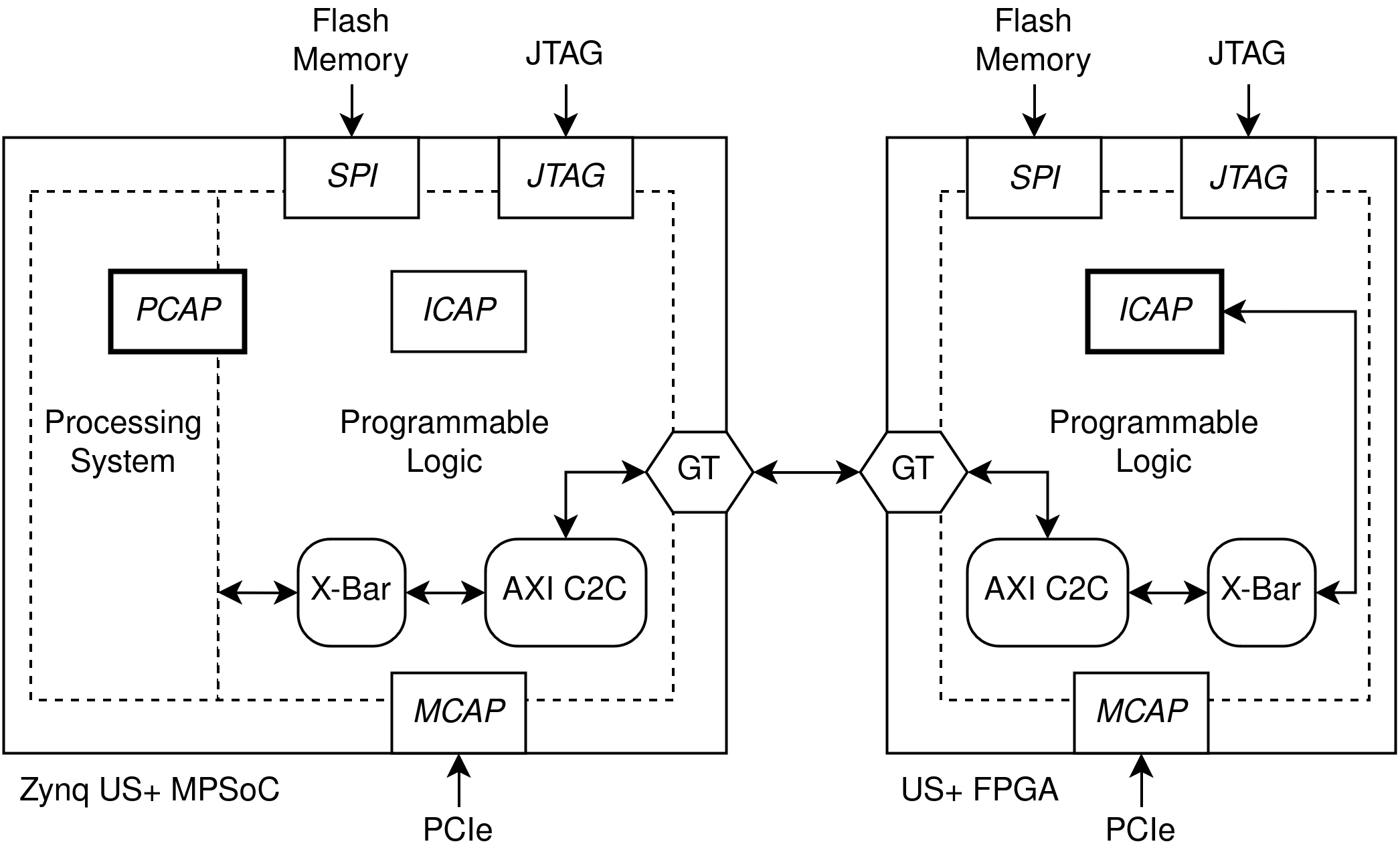}
    \caption{Resource extension of an \ac{mpsoc} using a peripheral \ac{fpga} connected via \ac{axi} \ac{c2c}, utilizing \acp{gt} on both devices. For both the \ac{pl} of the \ac{mpsoc} and for the \ac{fpga}, the most important configuration interfaces are shown. The interfaces that are to be used in the multi-device QiController are highlighted (\Ac{pcap} to configure the \ac{pl} in the \ac{mpsoc}, \Ac{icap} to configure the peripheral \ac{fpga}).}
    \label{fig:plExtension}
\end{figure}

Using \ac{axi} \ac{c2c} to expand the logic and interface resources of an \ac{mpsoc} by connecting an \ac{fpga} is a widely used approach. It is not only planned for the multi-device QiController; it is also already implemented in the aforementioned \ac{atca} blades developed at CERN, for example. The concept is shown in \autoref{fig:plExtension}. Since the connection is established between the \ac{pl} and the \ac{fpga}, it does not make a difference for the \ac{ps} of the \ac{mpsoc} if the resources are in the \ac{pl} of the \ac{mpsoc} itself or in the external \ac{fpga}. All resources are accessible from the same \ac{ps} interface, and the connection is transparent.

\autoref{fig:plExtension} also shows the most important configuration interfaces of the \ac{mpsoc}'s \ac{pl} and of an \ac{fpga} from AMD Xilinx. It can be seen that the \acf{icap} interface is the only one of the \ac{fpga} that is accessible from the \ac{ps} of the \ac{mpsoc} via \ac{axi} \ac{c2c} and without additional connections between the devices or loopbacks on the \ac{fpga}. The \ac{icap} interface is part of the \ac{fpga} and allows \acl{pr} from within the fabric. It must always be combined with a suitable IP core that provides control logic. All \acp{fpga} of the UltraScale and UltraScale+ family from AMD Xilinx offer this interface, as do all their \ac{mpsoc} devices. According to the manufacturer, the \ac{icap} interface is also the fastest option for \acl{pr} \cite{XAPP1338}. This is especially relevant as the configuration interface is usually the most crucial element when it comes to rapid \acl{pr}. The theoretical maximum throughput of the \ac{icap} interface is 800\,MB/s, which can also be achieved in practice under certain conditions and if the interface is overclocked from 100\,MHz to 200\,MHz \cite{dicarlo2015, duhem2011, pham2020}. This is much faster than other interfaces, such as the \ac{mcap} targeting \ac{pcie} with a bandwidth of typically 3 to 6\,MB/s or the \acf{pcap} intended for configuring the \ac{pl} from the \ac{ps} in an \ac{mpsoc} and achieving a maximum throughput of 256\,MB/s \cite{pham2020}. However, these values refer purely to the performance of the hardware and the \ac{fpga} firmware and do not include any software overhead, such as control from an operating system.

\section{Partial Reconfiguration}
\label{sec:partialReconfiguration}

During the design process, the \ac{fpga} is divided into a static region and one or more \aclp{rp}. One or more modules can be implemented for each \acl{rp}, which can later be exchanged dynamically at runtime. This corresponds to a time-division multiplexing of the hardware resources. To prevent signals from being unintentionally emitted or received by the \acl{rp} during \acl{pr}, the interfaces between the static design and the \acl{rp} should be disconnected in an orderly manner for the duration of the reconfiguration. This is typically accomplished using so-called decoupling logic.

Both major \ac{fpga} manufacturers Intel Altera and AMD Xilinx offer devices with \acl{pr} features \cite{intel_ug20179, xlnx_ug909}. However, the QiController is based on an \ac{rfsoc} device from AMD Xilinx, so this contribution only targets their UltraScale+ family of \ac{fpga} and \ac{mpsoc} devices. For the sake of clarity, the term \acl{pr} is used throughout this document, even though AMD Xilinx calls this feature \ac{dfx}.

The \ac{fpga} subsystem in Linux provides a vendor agnostic way for full and partial reconfiguration of \acp{fpga} \cite{linux_fpga_subsys}. It comprises two kinds of Kernel drivers: \Ac{fpga} managers and \ac{fpga} bridges. \Ac{fpga} managers implement one specific method to configure an \ac{fpga}. To accomplish this, they control all required hardware and \ac{fpga} firmware. \Ac{fpga} bridges control the decoupling logic to disconnect a \acl{rp} from its surroundings. One \ac{fpga} manager and any number of \ac{fpga} bridges are grouped in \ac{fpga} regions, which embody either one \acl{rp} or a full \ac{fpga}. \Ac{fpga} managers, \ac{fpga} bridges, \ac{fpga} regions, and their dependencies must be declared in the Linux device tree, which is a standardised form of describing the hardware components of an embedded device or computer. The device tree is evaluated by the Kernel at boot time and later on to initialize, manage and use these components. Device tree overlays provide a method to patch the device tree at runtime. They can be used to notify the Kernel about newly connected hardware or to unregister hardware before it is physically removed. In the same way they can also be used to inform the Kernel that logic is loaded or unloaded from a \acl{rp} or from the entire \ac{fpga}. Device tree overlays are even capable to actively load and unload logic when they are applied or removed.

\section{The Proposed Approach}
\label{sec:ourApproach}

In the architecture described in \autoref{fig:multiBoardPlatform} one \ac{rfsoc} acts as the central point for user interaction and updates via the network interface. This is referred to as the central \ac{rfsoc}. However, to optimise resource utilisation and maintain system modularity and expandability, each \ac{rfsoc} is responsible for managing, initialising, and updating the \acp{fpga} connected to it. As updates are merely reinitialisations that are carried out at runtime, and therefore do no differ from initialisations, the term initialisation is used for both operations in the following. The \ac{atca} standard prescribes a sophisticated \ac{hpm} infrastructure on every ATCA card which includes a low-bandwidth reconfiguration ability, however it is not suitable for fast \acl{pr} in real-time \cite{hpm, Calligaris2024}. The limited backplane connections prevent the implementation of a dedicated configuration connection for each associated \ac{fpga} board in the crate. Since the \ac{pcap} interface is solely available on \acp{mpsoc}, the only remaining option to enable the \acp{rfsoc} to manage the peripheral \acp{fpga} is to access their \ac{icap} interface through the \ac{axi} \ac{c2c} connection. This eliminates the need for a dedicated physical configuration connection. If the reconfiguration capability is used primarily for initialisation and not for rapidly multiplexing the \ac{fpga} fabric, this is also a particularly efficient solution, as the \ac{axi} \ac{c2c} connection is not required for other data transmission during initialisation. To enable the \ac{axi} \ac{c2c} connection and access to the \ac{icap} interface before initialisation, the \ac{fpga} automatically loads a small static design from a local flash memory at power-up. Because this static design only needs to include the \ac{axi} \ac{c2c} interface, the \ac{icap} controller, and decoupling logic, it can be kept independent of the application of the system, which is why it is reasonable to assume that it will rarely need to be updated.

As shown in \autoref{fig:multiBoardPlatform}, all peripheral \acp{fpga} in the system serve the same purpose, which is to interface \acp{dac} and increase the number of available analogue channels. Therefore, the same configuration file should be used on all of them to reduce the building and maintenance effort and to simplify scaling the system. However, it is not easily possible to reuse the same file for every \ac{fpga}, as \ac{axi} is an address-based bus that requires a unique address per node. To use the same configuration file on all peripheral \acp{fpga}, the \ac{axi} address space of the overall system must be subdivided. The most significant byte addresses an \ac{fpga} or the \ac{pl} of an \ac{rfsoc}, while the less significant four bytes are used for internal device addressing. This allows to truncate the device-specific part of the address before it is transferred via \ac{axi} \ac{c2c} to a peripheral \ac{fpga} \cite{krause2023_thesis}. Thus, the \acp{rfsoc} have 40-bit addressing, while the \acp{fpga} only have 32-bit addressing.

The initialisation of the entire system is controlled by the Linux operating system on the \acp{rfsoc}. For this purpose, the capabilities of the \ac{fpga} subsystem in Linux are used in combination with device tree overlays. In a three-stage procedure, device tree overlays are first applied on each \ac{rfsoc} independently to initialise the respective \ac{pl}. After that, device tree overlays are loaded to reveal the \acp{rfsoc} to each other and the static part of the peripheral \acp{fpga} to the respective Kernels on the \acp{rfsoc}. Finally, the \aclp{rp} of the peripheral \acp{fpga} are actively configured and introduced to the responsible Kernel with the third layer of device tree overlays. This procedure enables the system to be scaled by adding more \acp{rfsoc} and \acp{fpga} at runtime. The concrete hardware implementation of the overall system, composed of several chips, can thereby be incorporated in \ac{fpga} manager and \ac{fpga} bridge drivers.

\section{Evaluation Setups}
\label{sec:evaluationSetups}

Three different setups based on commercial evaluation cards were used to evaluate the functionality and performance of the approach. Each setup contains one AMD Xilinx ZCU102 to emulate an \ac{rfsoc} of the architecture in \autoref{fig:multiBoardPlatform}. Even though the ZCU102's \ac{mpsoc} is missing \acp{adc} and \acp{dac}, it is based on the same architecture as an \ac{rfsoc} and offers all relevant features for \acl{pr}. The peripheral \acp{fpga} in \autoref{fig:multiBoardPlatform} are emulated with AMD Xilinx VCU118 evaluation cards. One lane of an QSFP to SFP+ cable is used to realise the \ac{axi} \ac{c2c} connection between the boards.

\begin{figure}[b]
    \centering
    \includegraphics[width=3.0in]{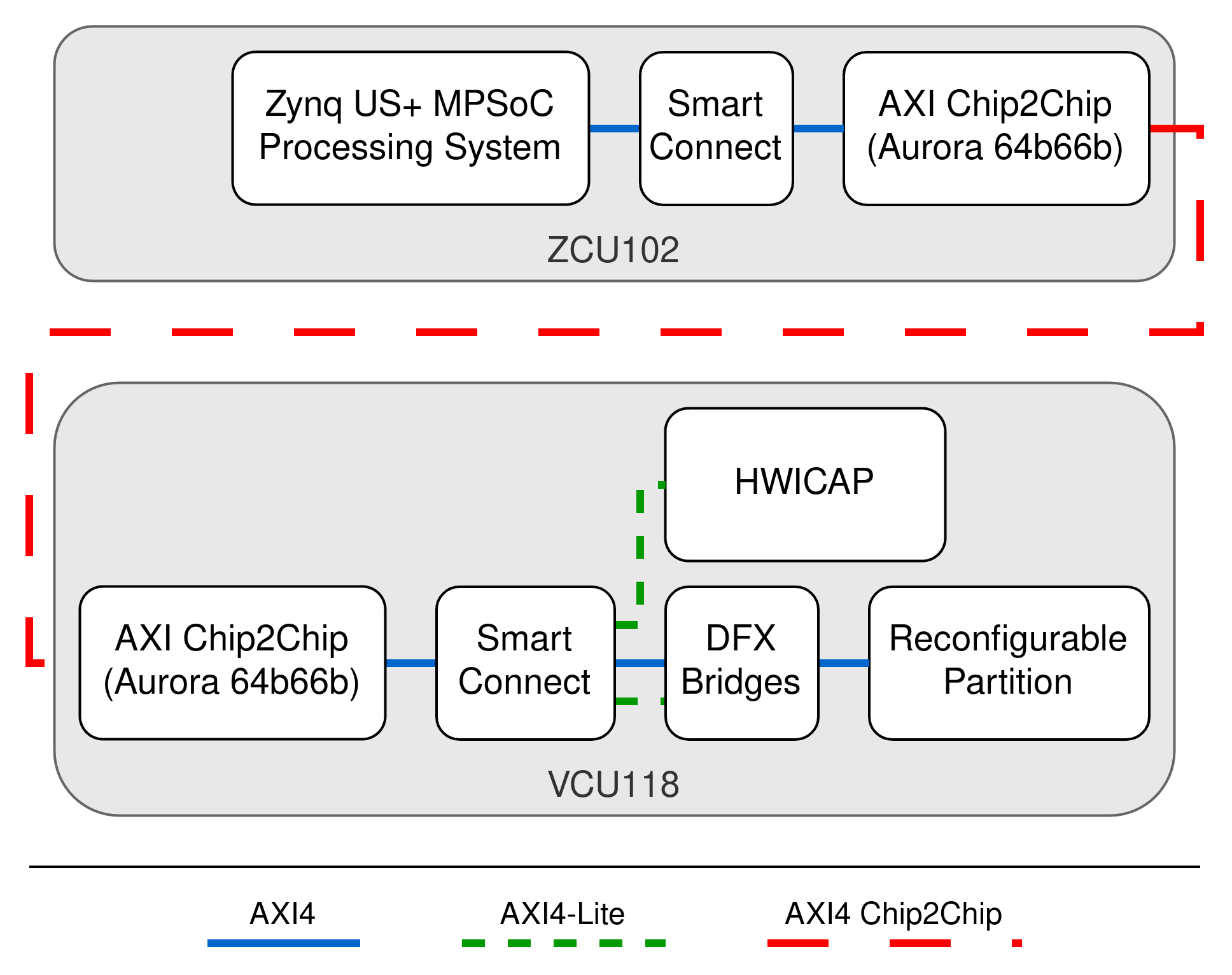}
    \caption{Setup with one AMD Xilinx ZCU102 and one AMD Xilinx VCU118 to test cross-chip \acl{pr} with a basic \ac{fpga} firmware architecture.}
    \label{fig:setupHWICAPC2C}
\end{figure}

To test the concept of \acl{pr} via \ac{axi} \ac{c2c}, the setup with one ZCU102 and one VCU118 shown in \autoref{fig:setupHWICAPC2C} was used \cite{krause2023_thesis}. The \ac{fpga} on the VCU118 hosts one instance of the \ac{axi} \ac{hwicap} IP core to access the \ac{icap} configuration interface of the \ac{fpga} and one instance each of an \ac{dfx} Decoupler and an \ac{dfx} \ac{axi} Shutdown Manager, which are \ac{fpga} bridges provided by the manufacturer. All of this connects to the \ac{ps} of the ZynqMP on the ZCU102 via \ac{axi} \ac{c2c} Bridge and Aurora 64B66B IP Cores on both devices. All IP cores used in this setup are provided by AMD Xilinx. To integrate the setup with the Linux \ac{fpga} subsystem, we have developed a custom FPGA Manager driver for the \ac{hwicap}, based on the example character device driver for MicroBlaze \cite{hwicap_microblaze}. 

\begin{figure}[b!]
    \centering
    \includegraphics[width=3.0in]{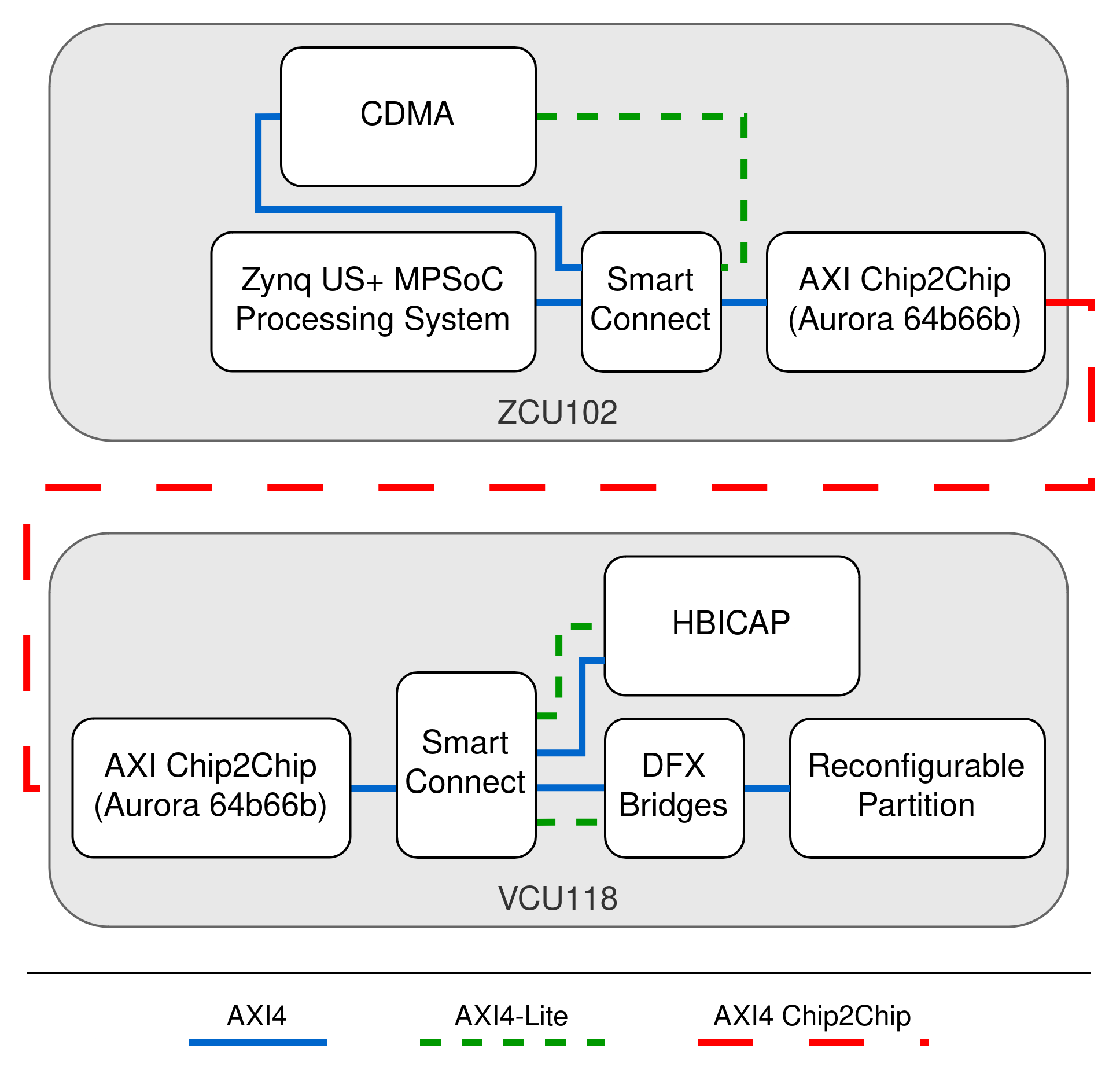}
    \caption{Setup with one AMD Xilinx ZCU102 and one AMD Xilinx VCU118 to test the performance of cross-chip \acl{pr} via a full AXI4 interface. The \ac{hbicap} IP core used features an AXI4-Lite control interface and an AXI4 data interface.}
    \label{fig:setupHBICAPC2C}
\end{figure}

\begin{figure}[b!]
    \centering
    \includegraphics[width=3.0in]{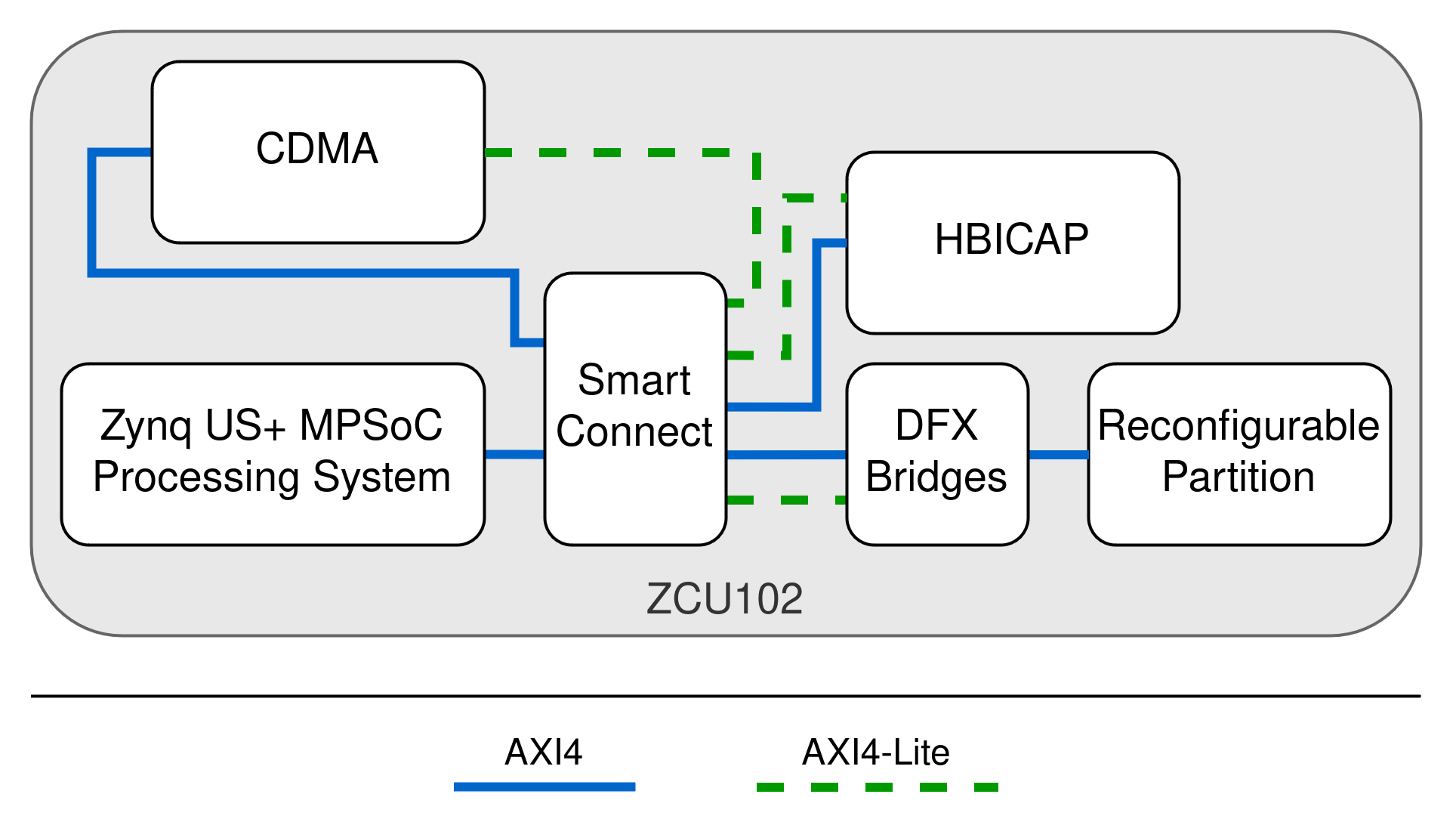}
    \caption{Setup with one AMD Xilinx ZCU102 to test the performance of the combination of \ac{hbicap} and \ac{cdma} without \ac{axi} \ac{c2c}.}
    \label{fig:setupHBICAP}
\end{figure}

The \ac{hwicap} IP core is a light weight and relatively easy to use way to access the \ac{icap} configuration interface. However, it only provides an AXI4-Lite data interface for transferring configuration data, which offers lower performance compared to a full AXI4 interface. To access the \ac{icap} interface with a full AXI4 connection, AMD Xilinx provides the more powerful and flexible \ac{axi} \acf{hbicap} IP core. \autoref{fig:setupHBICAPC2C} shows the adapted architecture that was used to explore the advantages of this IP core. The architecture also comprises a \acf{cdma} IP core on the \ac{mpsoc} to reduce processor load and exploit the full potential of the AXI4 interface. To control both the \ac{hbicap} and the \ac{cdma} IP core in this distributed configuration, we have again developed a custom \ac{fpga} manager driver. Finally, this setup was also tested with two VCU118 evaluation cards to better represent the architecture described in \autoref{fig:multiBoardPlatform}.

To evaluate the impact of the \ac{axi} \ac{c2c} connection on the performance of \acl{pr}, an equivalent setup without \ac{axi} \ac{c2c} was created as well. The layout is shown in \autoref{fig:setupHBICAP}. This setup is also more comparable to setups commonly used for the development and benchmarking of custom high-performance \ac{icap} solutions such as ZyCAP \cite{vipin2014}. This in turn makes it easier to compare the combination of \ac{hbicap} and \ac{cdma} with these high-performance solutions to assess whether they should also be tested in an \ac{axi} \ac{c2c} configuration.

To support the use of the \ac{hbicap} and \ac{hwicap} IP cores in the community, we have released our implementation of the corresponding \ac{fpga} manager drivers to the public \cite{cc_dfx_drivers}.

\section{Analysis}
\label{sec:analysis}

Several tests were conducted to quantify the suitability of the proposed approach for the initialisation of peripheral \acp{fpga} in the multi-device QiController. All measurements were performed using a system clock of 100\,MHz, and the \ac{icap} interface was not overclocked. Additionally, the line rate for the \ac{axi} \ac{c2c} connection was set to 10.3125\,Gbps. \ac{cpu} intensive startup tasks in the Linux userspace, such as the jitter-based initialisation of \textit{rng-tools}, were disabled, and tests were automatically executed once the \ac{axi} \ac{c2c} link was up, or, in the setup  without \ac{axi} \ac{c2c}, once Linux was fully booted. The reconfiguration time was determined in the \ac{fpga} fabric using a counter incremented with the system clock while the \ac{fpga} bridges were disconnected, measuring the time during which the \acl{rp} was not usable due to reconfiguration. The \ac{dfx} \ac{axi} Shutdown Manager initiated and stopped the counter. Each measurement was repeated ten times, separated by a one-second pause. After this, the system was hard rebooted and the entire procedure was repeated ten times, resulting in a total of 100 data points.

\begin{figure}[b]
    \centering
    \includegraphics[width=3.5in]{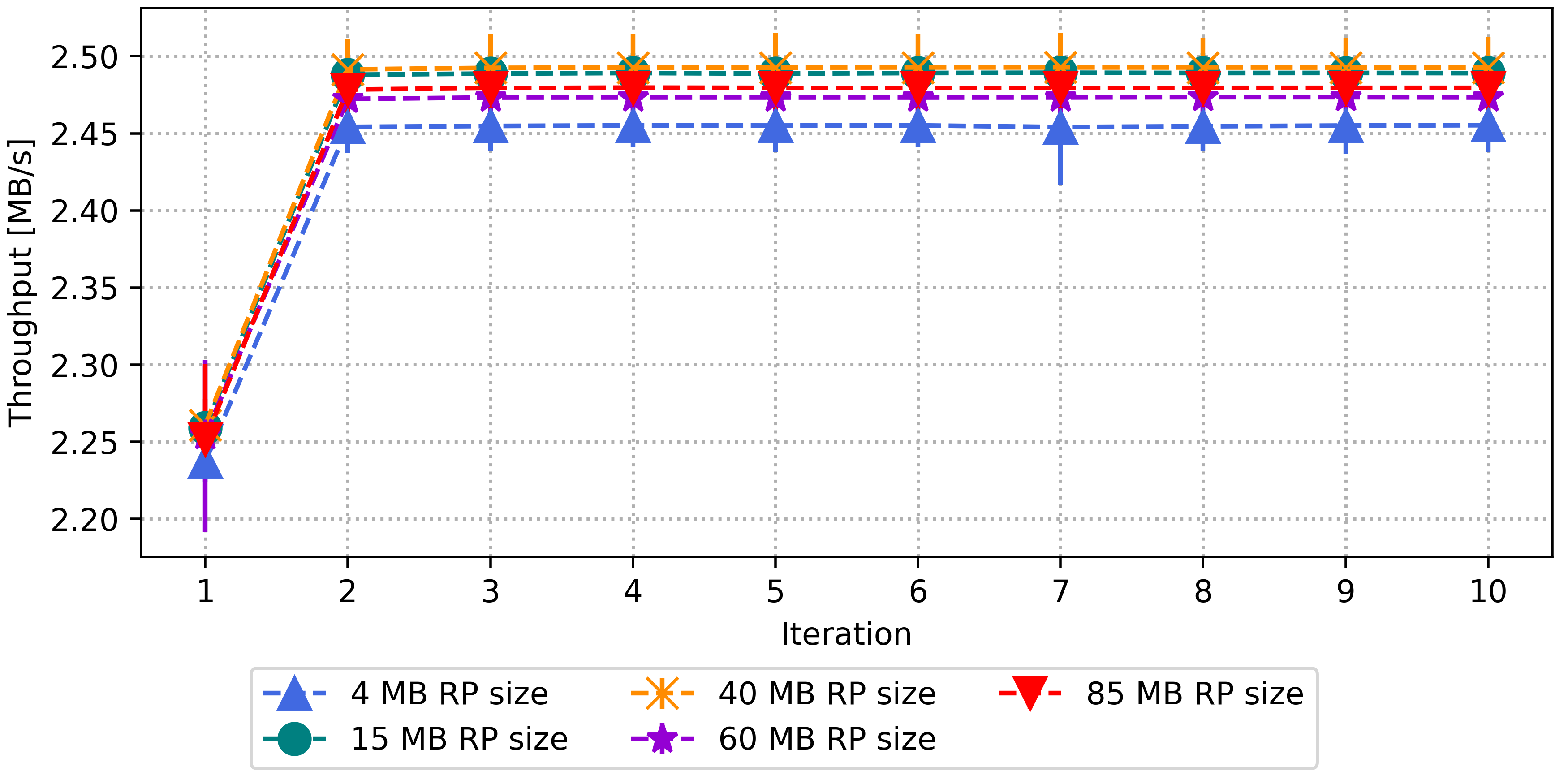}
    \caption{Arithmetically averaged throughput during \acl{pr} for several consecutive measurements performed directly after system startup and for differently sized \aclp{rp}. The consecutive measurements are separated by a one-second pause. Deviations are indicated by error bars. To make them more visible, the error bars were multiplied by a factor of five and therefore do not correspond to the scaling of the Y-axis. The presented data shows the behaviour of the setup with a \ac{hwicap} + \ac{axi} \ac{c2c} (\autoref{fig:setupHWICAPC2C}).}
    \label{fig:lpTrendHwicapC2c}
\end{figure}

\begin{figure}[t]
    \centering
    \includegraphics[width=3.5in]{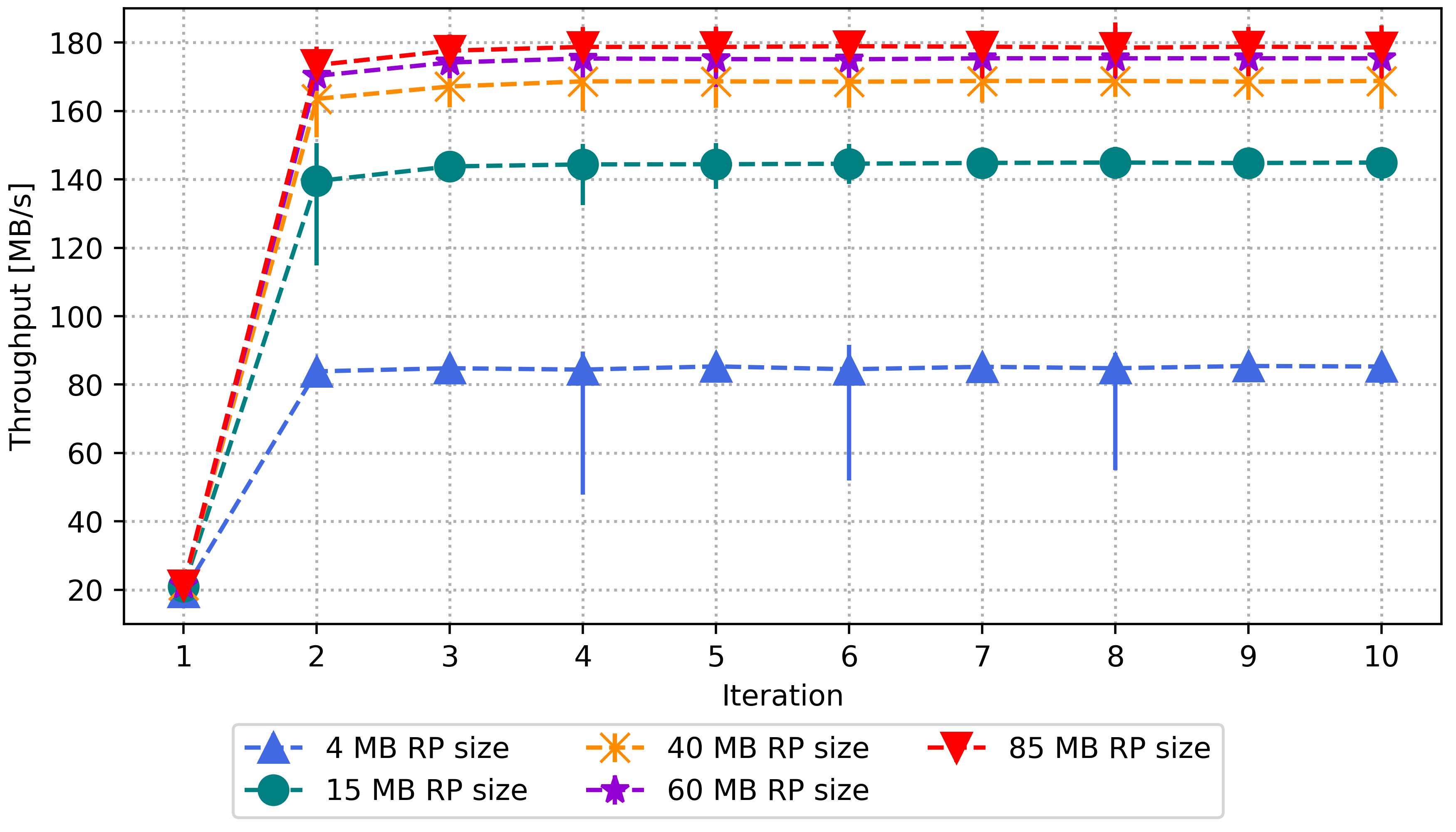}
    \caption{Arithmetically averaged throughput during \acl{pr} for several consecutive measurements performed directly after system startup and for differently sized \aclp{rp}. The consecutive measurements are separated by a one-second pause. Deviations are indicated by error bars. To make them more visible, the error bars were multiplied by a factor of five and therefore do not correspond to the scaling of the Y-axis. The presented data shows the behaviour of the setup with a \ac{hbicap} + \ac{axi} \ac{c2c} (\autoref{fig:setupHBICAPC2C}).}
    \label{fig:lpTrendHbicapC2c}
\end{figure}

\begin{figure}[t]
    \centering
    \includegraphics[width=3.5in]{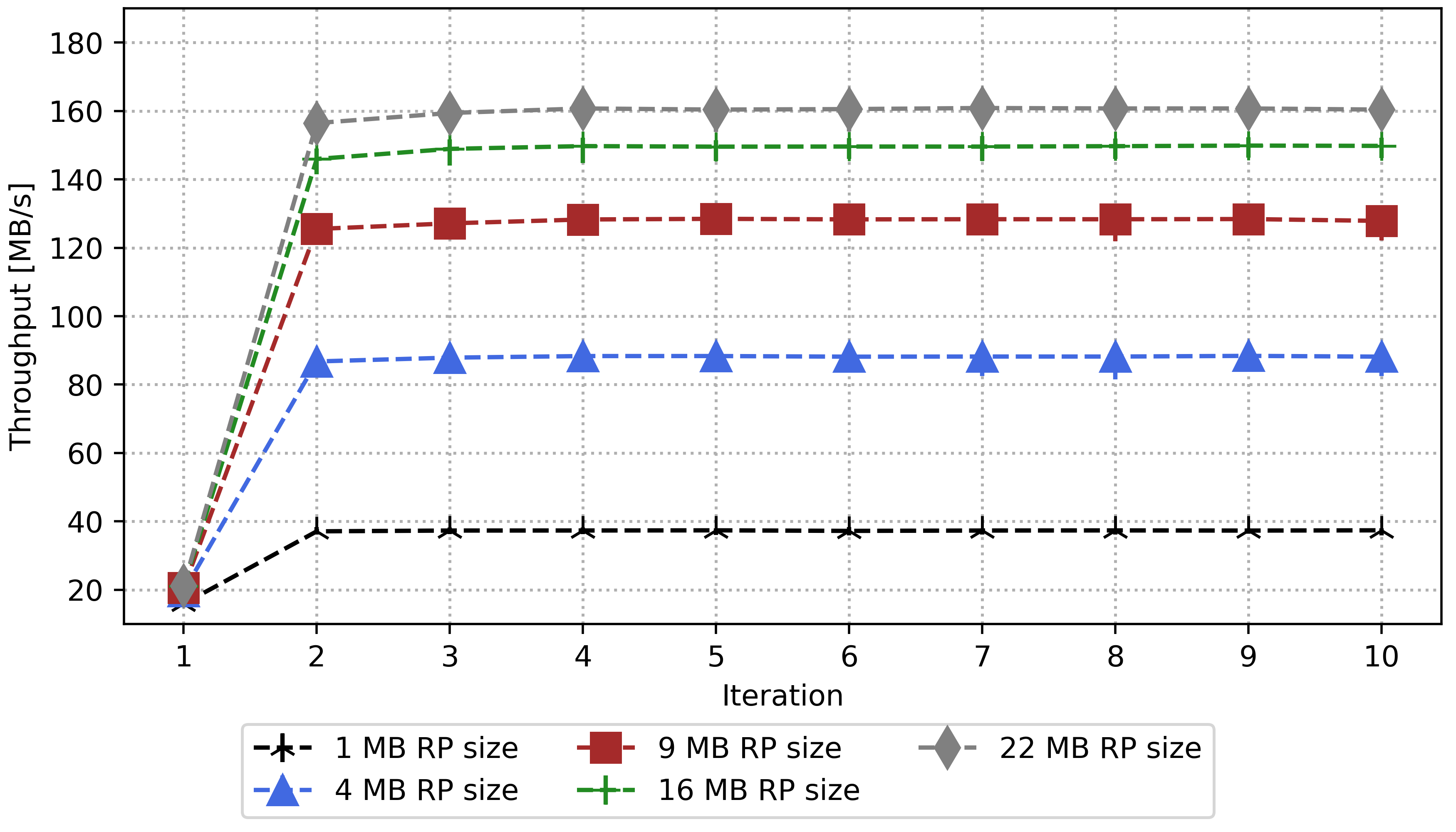}
    \caption{Arithmetically averaged throughput during \acl{pr} for several consecutive measurements performed directly after system startup and for differently sized \aclp{rp}. The consecutive measurements are separated by a one-second pause. Deviations are indicated by error bars. To make them more visible, the error bars were multiplied by a factor of five and therefore do not correspond to the scaling of the Y-axis. The presented data shows the behaviour of the setup with a \ac{hbicap} on a single ZCU102 (\autoref{fig:setupHBICAP}).}
    \label{fig:lpTrendHbicap}
\end{figure}

Figures \ref{fig:lpTrendHwicapC2c}, \ref{fig:lpTrendHbicapC2c}, and \ref{fig:lpTrendHbicap} provide an overview of the acquired measurement data. The overall highest reconfiguration throughput can be seen in \autoref{fig:lpTrendHbicapC2c} at 178\,MB/s and was achieved with the largest \acl{rp} and the \ac{axi} \ac{hbicap}. This corresponds to configuring 90\% of the resources of an AMD Xilinx XCVU9P in 0.5\,s and is 45\% of the maximum throughput that can be achieved using the 32-bit AXI4 bus at a clock rate of 100\.MHz. In contrast, the highest reconfiguration throughput with the \ac{axi} \ac{hwicap} is about two orders of magnitude lower at 2.5\,MB/s, as can be seen in \autoref{fig:lpTrendHwicapC2c}. All three Figures \ref{fig:lpTrendHwicapC2c}, \ref{fig:lpTrendHbicapC2c}, and \ref{fig:lpTrendHbicap} show minor fluctuations in the measurements and that the reconfiguration speed clearly depends on the size of the \acl{rp}. This is particularly visible in \autoref{fig:lpTrendHbicapC2c} and \autoref{fig:lpTrendHbicap}, showing results for the \ac{hbicap}. One exception are the first reconfigurations after the power-up of the system. All three figures indicate that in this case the reconfiguration throughput is significantly lower and does not depend on the size of the \acl{rp}. Figures \ref{fig:lpTrendHwicapC2c}, \ref{fig:lpTrendHbicapC2c}, and \ref{fig:lpTrendHbicap} show that all reconfigurations carried out directly after the start of the system are affected without exception. But only these, from the second reconfiguration onwards, the throughput is significantly higher and approximately constant across all following reconfiguration runs. A more detailed discussion of this effect can be found in \autoref{sec:resdis}.

\begin{figure}[t]
    \centering
    \includegraphics[width=3.5in]{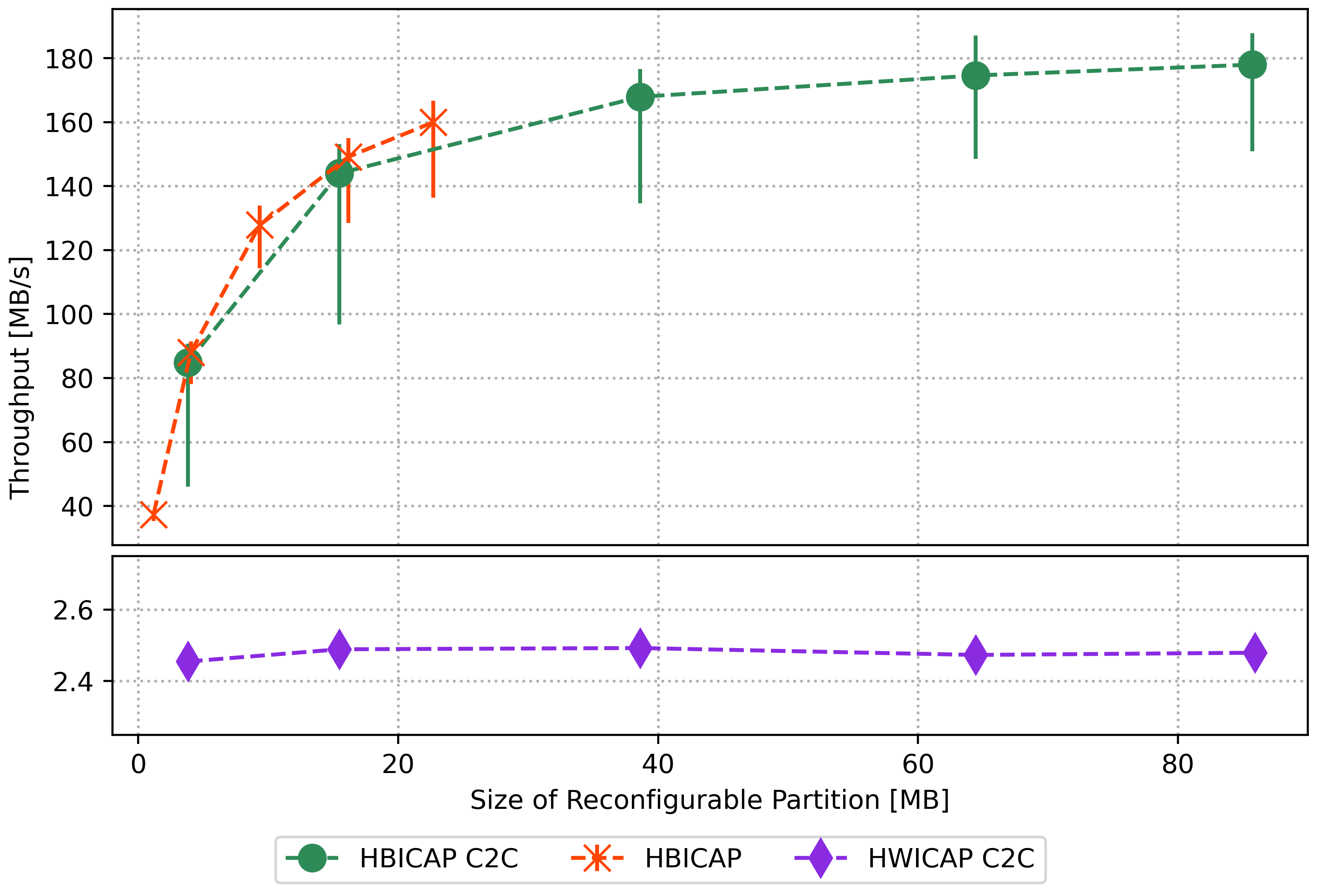}
    \caption{Arithmetically averaged throughput during \acl{pr} in relation to the size of the \acl{rp}. Deviations are indicated by error bars. To make them more visible, the error bars were multiplied by a factor of five and therefore do not correspond to the scaling of the Y-axis.}
    \label{fig:lpFileSize}
\end{figure}

\autoref{fig:lpFileSize} illustrates how the size of a \acl{rp} affects throughput during reconfiguration. This graph only contains values from the second reconfiguration onward to omit the above-mentioned effect of slow initial reconfiguration. The large gap of more than an order of magnitude between \ac{hbicap} and \ac{hwicap} based setups clearly shows that AXI4-Lite severely limits performance. This is mainly because \ac{axi} is a burst-based protocol and the maximum burst length for AXI4-Lite is limited to one data transfer per read and write request \cite{arm_ihi0022h}. In contrast, a full AXI4 interface supports bursts of up to 256 data transfers per request. This difference has an even greater impact on performance when an \ac{axi} \ac{c2c} connection is involved, since each request requires a handshake procedure, which takes longer when two independent chips are involved. The measurement results of the \ac{hbicap}-based systems show that in this case the size of the \acl{rp} has a significant influence on the throughput during reconfiguration. This observation is independent of whether \ac{axi} \ac{c2c} is used or not. Thus, the setup with \ac{axi} \ac{c2c} achieves even a higher maximum throughput than the single-board implementation on the ZCU102, because the \ac{fpga} on the VCU118 has significantly more resources than the \ac{pl} on the ZCU102, allowing for a larger \acl{rp}. In case of the ZCU102, more than 75\% of all resources where assigned to one \acl{rp} and with the VCU118 even more than 90\% were possible without any issues. In case of the \ac{hwicap} based setup, there is no dependence of the reconfiguration throughput on the partition size.

In all measurement setups, the content of the \acl{rp} and the order in which the configurations were applied had no effect on the reconfiguration speed. Therefore, this will not be discussed further.

\section{Results and Discussion}
\label{sec:resdis}

This section will discuss several inferences drawn from the preceding diagrams. \autoref{fig:lpFileSize} shows that reconfiguration with the \ac{hwicap} IP core via \ac{axi} \ac{c2c} is in some cases nearly two orders of magnitude slower than with the \ac{hbicap} IP core. In case of the \ac{hwicap}, the reconfiguration throughput is limited by \ac{axi} \ac{c2c}, as can be seen from the corresponding graph in \autoref{fig:lpFileSize} which shows no dependence between the size of the \acl{rp} and the throughput. This is also underlined by the fact that in other comparable studies without \ac{axi} \ac{c2c}, the \ac{hwicap} is significantly faster and the performance gap to the \ac{hbicap} is considerably smaller \cite{vipin2014, pham2020}. In contrast, reconfiguration with the \ac{axi} \ac{hbicap} using full AXI4 for data transfer is not limited by \ac{axi} \ac{c2c}. This can be seen from the corresponding graphs in \autoref{fig:lpFileSize}, which clearly show the dependence of the throughput on the size of the \acl{rp}. From \autoref{fig:lpFileSize} it can also be seen that the impact of \ac{axi} \ac{c2c} is almost negligible when the \ac{hbicap} is used. This is especially visible at the 4\,MB data point, where the setup with \ac{axi} \ac{c2c} achieves an average of 84.8\,MB/s versus 88.0\,MB/s in the setup without \ac{axi} \ac{c2c}, corresponding to a marginal deviation of approximately 3.4\%.

The slow first reconfiguration as seen in Figures \ref{fig:lpTrendHwicapC2c}, \ref{fig:lpTrendHbicapC2c}, and \ref{fig:lpTrendHbicap} was also investigated. This effect is only observed after a hard restart of the \ac{mpsoc}, and not when the peripheral \ac{fpga} is restarted. Thus, it is unlikely that this effect is related to \ac{axi} \ac{c2c}, the \ac{icap} interface controller, the \ac{fpga} bridges, or the \acl{rp} itself. Since the effect occurs with both the \ac{hbicap} and \ac{hwicap}, it is also not related to the \ac{cdma}. Furthermore, the effect also occurred if the first measurement was not started immediately after power-up, but with a delay in the order of minutes. It also made no difference whether the same or different bitsteams were loaded one after the other, which is why the phenomenon cannot be attributed to the caching of this date. Most likely, this effect is caused by the \ac{ps} of the \ac{mpsoc} or by the Linux operating system running on it. However, the exact cause could not be identified yet.

As was previously shown by Vipin and Fahmy, with a custom controller the \ac{icap} interface can reach a maximum throughput of 382\,MB/s, without overclocking the configuration interface and utilising a system clock of 100\,MHz \cite{vipin2014}. This is approximately twice as fast as the highest average reconfiguration throughput we have achieved with the \ac{hbicap} from AMD Xilinx at 178\,MB/s (see \autoref{fig:lpFileSize}). Based on our investigations so far, it is not possible to give a statement about how much of this deviation is caused by the \ac{hbicap} IP core, the Linux operating system, or our custom \ac{hbicap} \ac{fpga} manager. Therefore, we intend to continue our work by investigating the performance of a custom \acs{icap} controller in the \ac{c2c} configuration. Additionally, further improvements may also be achieved through optimised settings in Linux, for example with regard to scheduling, caching or the initialization of drivers.  There is presumably still room for improvement in the custom \ac{hbicap} \ac{fpga} manager too. In particular, possibilities to accelerate the first reconfiguration of a \acl{rp} after system startup should be investigated, because this is crucial if the proposed approach is eventually being used to initialize the system. Nonetheless, the reconfiguration speed achieved in this work, and the proposed method utilising the \ac{axi} \ac{hbicap} IP core are already suitable for the multi-device QiController, because the peripheral \acp{fpga} are only initialized once with \acl{pr} during the startup procedure of the instrument and occasionally to perform system updates at runtime. A reconfiguration time in the order of seconds meets the requirements in both cases.

\section{Conclusion}

This work presented an \ac{axi} \ac{c2c}-based approach to initialise modular and scalable heterogeneous \ac{daq} systems composed of multiple \acp{mpsoc} and \acp{fpga} from AMD Xilinx. It uses the Linux operating system on the \ac{mpsoc} to manage the entire process. In particular, the \ac{fpga} subsystem of Linux is used together with device tree overlays to perform partial reconfiguration on the peripheral \acp{fpga}. Our implementation of suitable \ac{fpga} manager drivers is released to the public \cite{cc_dfx_drivers}. A series of measurements were performed on three different setups using evaluation cards to determine the performance and reliability of the method.

A firmware architecture made entirely of AMD Xilinx IP cores and built around the combination of \ac{axi} \ac{hbicap} and \ac{axi} \ac{cdma} proved the feasibility of using this architecture for the multi-device QiController, currently being designed following the ATCA standard. The reconfiguration throughput of 178\,MB/s achieved with the current implementation has the potential to even further improve with the use of custom \ac{icap} and \ac{cdma} controllers. Therefore, we intend to explore this avenue in the coming months.

\appendices

\end{document}

%% file: acronyms.tex
\DeclareAcronym{soc}{
short = SoC ,
long = system-on-chip ,
plural = s
}
\DeclareAcronym{mpsoc}{
short = MPSoC ,
long = multiprocessor system-on-chip ,
plural = s
}
\DeclareAcronym{rfsoc}{
short = RFSoC ,
long = radio-frequency system-on-chip ,
plural = s
}
\DeclareAcronym{fpga}{
short = FPGA ,
long = field-programmable gate array ,
plural = s
}
\DeclareAcronym{api}{
short = API ,
long = application programming interface ,
plural = s
}
\DeclareAcronym{sdr}{
short = SDR ,
long = software-defined radio ,
plural = s
}
\DeclareAcronym{cpu}{
short = CPU ,
long = central processing unit ,
plural = s
}
\DeclareAcronym{gpu}{
short = GPU ,
long = graphics processing unit ,
plural = s
}
\DeclareAcronym{ai}{
short = AI ,
long = artificial intelligence ,
plural = s
}
\DeclareAcronym{asic}{
short = ASIC ,
long = application-specific integrated circuit ,
plural = s
}
\DeclareAcronym{daq}{
short = DAQ ,
long = data acquisition ,
plural =
}
\DeclareAcronym{adc}{
short = ADC ,
long = analogue-to-digital converter ,
plural = s
}
\DeclareAcronym{dac}{
short = DAC ,
long = digital-to-analogue converter ,
plural = s
}
\DeclareAcronym{pl}{
short = PL ,
long = programmable logic ,
plural = s
}
\DeclareAcronym{ps}{
short = PS ,
long = processing system ,
plural = s
}
\DeclareAcronym{atca}{
short = ATCA ,
long = advanced telecommunications computing architecture ,
plural =
}
\DeclareAcronym{icap}{
short = ICAP ,
long = internal configuration access port ,
plural =
}
\DeclareAcronym{pcap}{
short = PCAP ,
long = processor configuration access port ,
plural =
}
\DeclareAcronym{mcap}{
short = MCAP ,
long = media configuration access port ,
plural =
}
\DeclareAcronym{hwicap}{
short = HWICAP ,
long = hardware internal configuration access port ,
plural =
}
\DeclareAcronym{hbicap}{
short = HBICAP ,
long = high bandwidth internal configuration access port ,
plural =
}
\DeclareAcronym{sfp}{
short = SFP ,
long = small form-factor pluggable ,
plural =
}
\DeclareAcronym{kit}{
short = KIT ,
long = Karlsruhe Institute of Technology ,
plural =
}
\DeclareAcronym{pcie}{
short = PCIe ,
long = Peripheral Component Interconnect Express ,
plural =
}
\DeclareAcronym{pr}{
short = PR ,
long = partial reconfiguration ,
plural =
}
\DeclareAcronym{aws}{
short = AWS ,
long = Amazon Web Services ,
plural =
}
\DeclareAcronym{dfx}{
short = DFX ,
long = Dynamic Function eXchange ,
plural =
}
\DeclareAcronym{cdma}{
short = CDMA ,
long = central direct memory access ,
plural =
}
\DeclareAcronym{rp}{
short = RP ,
long = reconfigurable partition ,
plural = s
}
\DeclareAcronym{c2c}{
short = C2C ,
long = Chip2Chip ,
plural =
}
\DeclareAcronym{fdm}{
short = FDM ,
long = frequency division multiplexing ,
plural =
}
\DeclareAcronym{axi}{
short = AXI ,
long = Advanced eXtensible Interface ,
plural =
}
\DeclareAcronym{gt}{
short = GT ,
long = gigabit transceiver ,
plural = s
}
\DeclareAcronym{hpm}{
short = HPM ,
long = hardware platform management ,
plural = 
}